\documentclass[aps,prl,twocolumn,groupedaddress,showpacs]{revtex4}
\usepackage{dcolumn}% Align table columns on decimal point
\usepackage{bm}% bold math

\begin{document}

\preprint{}

\title{Reducing the Leak Rate from a Damaged Oil Well by Filling It with Dense Streamlined Objects}

\author{Louis A. Bloomfield}
\email{lab3e@virginia.edu}
\affiliation{Department of Physics, University of Virginia, Charlottesville, 
Virginia 22904}

\date{\today}

\begin{abstract}
The enormous pressure lifting the column of oil in a leaking oil well can thwart efforts to seal the top of the well and prevent oil from rising. When the oil cannot be stopped completely, we propose to slow its flow by filling the well with a porous medium. That medium consists of countless small, dense, streamlined objects that are dropped into the well and descend through the rising oil at terminal velocity. The resulting heap of objects couples to the oil via viscous and drag forces, dissipating the oil's energy and upward momentum and significantly reducing its rate of flow.

\end{abstract}

\pacs{91.65.My,89.30.aj,47.85.Dh,47.85.L-}

\maketitle

The catastrophic oil leak that followed the destruction of the Deepwater Horizon drilling platform in the Gulf of Mexico highlights the need for new ways to limit the flow of oil from a crippled oil well. Since permanently stopping a major leak can take months, an interim solution that quickly reduces the leak rate could significantly mitigate environmental damage.

While the permanent solution is being implemented, we proposed to fill the leaking well with small, dense, streamlined objects that are dropped into the well and descend at terminal velocity through the rising oil. As they accumulate, these objects become a porous medium that impedes fluid flow in the well and reduces the leak rate.

Drilling an oil well is exercise in pressure management.\cite{lyons} Oil, water, and gas are highly buoyant in the earth's crust, and they can move upward rapidly given the opportunity. Keeping those fluids under control is a key challenge for petroleum engineering.

Since the density of earth's crust is approximately 2800 kg/${\mathrm m}^3$, the ambient pressure at the bottom of a 5-km oil well is roughly 140 MPa. Only 40 MPa are needed to support the weight of a 5-km column of crude oil (800 kg/${\mathrm m}^3$), so there is an excess bottom pressure of about 100 MPa (15,000 psi) acting to lift that column of oil upward. While that excess pressure is typically reduced by dynamic pressure drops in moving fluids and by the support of solid mineral formations, it can cause major problems in deep wells.

During drilling, a deep oil well is kept full of drilling mud---a heavy liquid with a carefully regulated density. Hydrostatic pressure in that mud balances the ambient pressure in the well and prevents unwanted fluid motion. If the mud density is too low, however, the excess bottom pressure can lift it and blow it out of the well; if the mud density is too high, its unnecessary weight can damage mineral formations near the bottom of the well. The narrow range of safe mud densities becomes even narrower as the well becomes deeper\cite{sylty} and mistakes or bad luck can lead to disaster.

If drilling mud escapes from the well, either through its top or through its walls\cite{Schneider}, low-density fluid can enter the well and begin to rise. Stopping that rising fluid requires an enormous top pressure and pushing it back down the well to reintroduce the dense mud requires still more pressure. If the top portion of the well is damaged or weak, those remedies may not be possible.

The standard technique for permanently stopping fluid flow in a seriously damaged well is to drill a relief well adjacent to the first and use that relief well to fill the damaged well from the bottom up with dense mud and/or cement.\cite{grace,flak,aadnoy} Working from the bottom of the damaged well avoids the need to pressurize its top. Drilling the new well, however, takes time and care because it is fraught with the same perils that may have caused the original well to fail.

While the relief well is being designed and constructed, efforts can be made to reduce the flow in the damaged well. But while there are many well-established techniques for increasing and improving that flow,\cite{porges} including hydraulic fracturing of the oil-bearing formations,\cite{guo} introduction of "proppant" particles into those formations to keep channels open\cite{guo,vanbatenburg}, and gravel packing to filter sand out of the fluid stream\cite{forrest,parlar}, techniques for deliberately impairing the flow are unexplored. Similarly, hydrodynamic studies of fluid flow in oil wells\cite{ribeiro,rodriquez,guo,porges} have focused on improving rather than reducing flow.

We propose to employ gravity and fluid dynamics to reduce the flow of oil in the damaged well. By filling the well with dense objects, drawn downward by gravity alone, we gradually increase the effective density of the fluid it contains. As the fluid's density increases, both its buoyancy and its flow rate decrease. When the fluid's effective density reaches the density of the earth's crust, the flow rate will have dwindled close to zero.

We assume that the well is a typical one; that its bottom is porous rock or sand rather than a vast underground cavern. The well thus has a well-defined bottom and a limited volume, making it possible to accumulate a heap of objects within the well and even to fill it completely.

On the length scale of the well hole, viscous and drag forces can couple the oil to those objects so that the hole is effectively filled with a high-density fluid. If iron objects (7874 kg/${\mathrm m}^3$) are used, the flow should essentially stop when iron occupies about 30\% of the well's volume. If lead objects (11340 kg/${\mathrm m}^3$) are used, only about 20\% of the well's volume needs to filled with lead.

The object-filled well resembles a porous medium. Oil flowing through the narrow channels in that medium dissipates ordered energy as its struggles against drag and viscous forces. The energy supplied to the oil by the enormous pressure beneath it is wasted as thermal energy and the upward momentum given to the oil by the bottom pressure is transferred to the objects and walls of the well.

The key challenge in this method is to ensure that the dense objects descend to the bottom of the well and accumulate there. Each object will descend only if its downward terminal velocity ${\mathbf v}_t$ relative to the fluid is greater in magnitude than the upward velocity of that fluid ${\mathbf v}_f$. This requirement is fairly easy to achieve when $|{\mathbf v}_f|$ is 5 m/s or less and is still within reach when $|{\mathbf v}_f|$ is 20 m/s.

An object falling through an upward moving fluid reaches terminal velocity when the three forces acting on it, namely its downward weight ${\mathbf W}$, the upward drag force ${\mathbf F}_D$, and the buoyant force ${\mathbf F}_B$, sum to zero:
\begin{equation}
{\mathbf W}+{\mathbf F}_D+{\mathbf F}_B=0.
\label{eq:one}
\end{equation}
The drag force can be written as:\cite{wegener87,anderson142185}
\begin{equation}
{\mathbf F}_D={\frac{1}{2}}\rho_f|{\mathbf v}|^2\hat {{\mathbf z}}C_DA,
\label{eq:two}
\end{equation}
where $\rho_f$ is the density of the fluid, ${\mathbf v}$ is the downward velocity of the object relative to the fluid, $C_D$ is the drag coefficient for the object in the fluid, and $A$ is the effective cross-sectional area of the object.

A spherical object of radius $r$ and density $\rho_s$ has a downward weight given by
\begin{equation}
{\mathbf W}=-{\frac{4}{3}}\pi r^3\rho_sg\hat{{\mathbf z}},
\label{eq:three}
\end{equation}
where $-g\hat{{\mathbf z}}$ is the acceleration due to gravity. That sphere experiences an upward buoyant force equal in magnitude to the weight of the fluid it displaces:
\begin{equation}
{\mathbf F}_B={\frac{4}{3}}\pi r^3\rho_fg\hat{{\mathbf z}}.
\label{eq:four}
\end{equation}
If we insert the cross-sectional area for a sphere ($A=\pi r^2$) in Eq. \ref{eq:two}, substitute Eqs. \ref{eq:two}, \ref{eq:three}, and \ref{eq:four} into Eq. \ref{eq:one}, set ${\mathbf v}={\mathbf v}_t$, and solve for $r$, we obtain the radius of the sphere as a function of its terminal velocity:
\begin{equation}
R={\frac{3}{8}}\frac{C_D|{\mathbf v}_t|^2}{g}\frac{\rho_f}{(\rho_s-\rho_f)}
\label{eq:five}
\end{equation}
A sphere is a hydrodynamically blunt object with a drag coefficient $C_D\simeq 2.1$ at Reynolds number $Re_d=30$ and $C_D\simeq 0.45$ at $Re_d=1000$.\cite{haider} Table \ref{tab:iron} lists approximate terminal velocities for iron spheres ($\rho_s = 7874$ kg/${\mathrm m}^3$) of various radii descending through crude oil ($\rho_f = 800$ kg/${\mathrm m}^3$, dynamic viscosity 0.05 Pa$\cdot$s). Even at deep-well pressures, the density of crude oil increases by at most a few percent and its viscosity typically remains well below 0.05 Pa$\cdot$s.\cite{ahrabi87,ahrabi89,schmidt} Moreover, the presence of natural gas in the oil, whether dissolved, gaseous, or supercritical, reduces the density and viscosity of the fluid and increases the terminal velocities of the objects.

\begin{table}[!ht]
\caption{Terminal Velocities of Iron Spheres and Teardrops in Crude Oil}
\begin{center}
\begin{ruledtabular}
\begin{tabular}{ccccc}
$|{\mathbf v}_t|$ (m/s)&$r_{sphere}$ (cm)&$Re_{d,sphere}$&$r_{drop}$ (cm)&$Re_{d,drop}$\\ \hline
0.5&0.2&30&&\\
1&0.4&130&&\\
2&0.9&600&&\\
3&1.6&1500&&\\
4&2.7&3500&0.42&530\\
4.5&3.5&5000&0.53&760\\
5&4.3&7000&0.65&1000\\
10&20&65000&2.6&8000\\
20&&&10&70000\\
\end{tabular}
\end{ruledtabular}
\end{center}
\label{tab:iron}
\end{table}

When the upward velocity of oil in the well hole is 5 m/s or less, fist-sized or smaller iron spheres will descend to the bottom of the well and accumulate there. Those initial spheres will reduce the upward velocity of the flow, allowing smaller spheres to follow. Eventually the flow will be so slow that even iron shot or sacks containing scrap iron and heavy mud will be able to descend into the well to curtail the flow.

For upward flows exceeding 5 m/s, however, iron spheres would have to be fairly large, and they might not be able to descend into narrow well bores (diameters less than about 25 cm), particular when drill pipe remains in the well. Instead, hydrodynamically streamlined objects are necessary. Elongating a sphere into a teardrop shape can reduce its drag coefficient to $C_D\simeq 0.12$ at $Re_d>100$,\cite{wegener104} while approximately doubling its weight and buoyancy. The terminal velocities of these streamlined teardrops are much greater than those of spheres, so smaller radii can be used (see Table \ref{tab:iron}).

At upward flow rates exceeding 10 m/s, more elongated hydrofoils with drag coefficients less than 0.1 will be needed to penetrate into the narrowest parts of the well. The design of these hydrofoils is beyond our expertise, but it is likely that carefully designed iron hydrofoils can descend into wells with upward flow rates exceeding 20 m/s.

The oil's velocity profile is not exactly uniform across the well's open bore. In regions of laminar flow we expect Poiseuille flow with its reduced velocities near stationary surfaces and increased velocities far from surfaces. Dense objects will therefore tend to descend near the walls of the well. In regions of turbulent flow, we expect a more uniform velocity profile.

The damaged Mississippi Canyon Block 252 well in the Gulf of Mexico is estimated to have been leaking at most 60,000 barrels/day and its narrowest steel casing has an internal radius of 8.9 cm. Had the well contained only oil, that oil's upward velocity would have been less than 4.5 m/s. As shown in Table \ref{tab:iron}, iron spheres of $\mathrm{radius} > 3.5$ cm and iron teardrop of $\mathrm{radius} > 0.53$ cm would have descended to the bottom of that well.

A drilling pipe, however, occupies some of the volume inside the well's casing. That pipe would have prevented iron spheres of $\mathrm{radius} > 3.5$ cm from descending all the way to the bottom of the well. But even if those spheres could not enter the narrowest of the well's several casings and could only fill, for example, the top half of the well, they would still have virtually stopped the oil flow. Iron teardrops of $\mathrm{radius} > 0.53$ cm would have descended all the way to the bottom in spite of the drill pipe. To the extent that the drilling pipe increased upward oil velocities in the well, some increase in the radii of the iron teardrops would have been required and iron or lead hydrofoils might even have been necessary.

As each object descends through the oil at terminal velocity, it extracts upward momentum from the oil at a rate equal to its weight and it transfers that momentum to the earth via gravity. When the object settles onto the heap at the bottom of the well, however, its velocity relative to the oil decreases and its weight is partly supported by the heap rather than by the oil. We should therefore expect the rate at which the stationary object extracts upward momentum from the oil to be less than its weight, possibly much less.

Fortunately, several effects work together to maintain a high momentum transfer rate. First, a streamlined object is unlikely to settle in its most streamlined orientation. Second, oil flowing through the narrow channels between settled objects has a low Reynolds number and experiences greatly enhanced viscous drag (skin friction).

Third, as the objects pile up in the bottom of the well, they press tightly against the well walls. Since the surfaces are well-lubricated (naturally!), friction plays a limited role. But the walls are not perfectly smooth, so objects can wedge themselves into nooks and crannies, particularly when there is drill pipe in the well. These wedged objects can convey large amounts of upward momentum from the oil to the well walls.

To maximize the upward momentum transfer from the oil to the objects to the earth, it is prudent to choose objects that only barely descend into the rising oil and that, if possible, are relatively small and streamlined. They will then form the desired porous medium, couple strongly to the oil, and grip the walls of the well securely.

As the oil flow slows, smaller objects will be able to descend and we envision gradually reducing the sizes of the objects fed into the well. At each moment, an assortment of object sizes might be dropped into the well, allowing the well to choose which it keeps and to reject those with inadequate terminal velocities. One could even pour a random assortment of old iron and steel hardware down the well hole, letting the well keep only those that manage to descend through the oil. Objects that can be streamlined easily, such as pieces of iron rebar with flattened tails to orient them in the flow, could be added to the mixture to strengthen the coupling between oil, objects, and walls.

To give the heap of iron objects additional grip on the walls of the well, sharp hardened steel objects could be included in the mixture. Their cutting edges would allow them to gouge into the steel well casing and cling to it strongly. Although expensive, sharp tungsten carbide objects would be even more effective at gripping the well walls and their extreme densities (15800 kg/${\mathrm m}^3$) would aid their descents into the well.

Iron and steel objects are particularly cost effective, but denser materials could be used if necessary to descend into fast rising oil. Silver bullets (10490 kg/${\mathrm m}^3$) come to mind. Of course, lead (11340 kg/${\mathrm m}^3$) would be much less expensive than silver and lead spheres and teardrops would have greater terminal velocities than iron objects of equal dimensions. Table \ref{tab:lead} lists approximate terminal velocities for lead spheres and teardrops ($\rho_s = 11340$ kg/${\mathrm m}^3$) in crude oil. For the estimated 4.5 m/s maximum upward velocity of oil in the Mississippi Canyon Block 252 well, lead spheres of $\mathrm{radius} > 2.3$ cm and lead teardrops of $\mathrm{radius} > 0.35$ cm would have descended to the bottom of the leaking Mississippi Canyon Block 252 well in the absence of drill pipe.

\begin{table}[!ht]
\caption{Terminal Velocities of Lead Spheres and Teardrops in Crude Oil}
\begin{center}
\begin{ruledtabular}
\begin{tabular}{ccccc}
$|{\mathbf v}_t|$ (m/s)&$r_{sphere}$ (cm)&$Re_{d,sphere}$&$r_{drop}$ (cm)&$Re_{d,drop}$\\ \hline
0.5&0.16&26&&\\
1&0.32&100&&\\
2&0.7&430&&\\
3&1.2&1100&&\\
4&1.8&2300&0.28&360\\
4.5&2.3&3300&0.35&500\\
5&2.9&4600&0.44&700\\
10&13.5&43000&1.7&5600\\
20&&&7&45000\\
\end{tabular}
\end{ruledtabular}
\end{center}
\label{tab:lead}
\end{table}

Lead is much softer than iron and its tendency to deform under stress has advantages and disadvantages. An advantage is that deformable lead objects will strengthen the coupling between the oil, objects, and walls as they deform under the weight of the objects above them. Lead objects may even deform to the point of forming a true seal within the bore of the well hole.

A disadvantage, however, is that if the lead objects deform significantly before they are deeply buried in the heap and wedged against the walls, they may be lifted upward by increased drag forces and pressure gradients and blown out of the well. However, a carefully chosen blend of iron, lead, tungsten carbide, and brittle iron (e.g., pig iron) objects might succeed in forming a seal that is deep enough in the object heap to withstand the pressure difference across it.

Even without a seal, the average density in the well could easily be raised above 5000 kg/${\mathrm m}^3$ for iron and above 7000 kg/${\mathrm m}^3$ for lead. With iron or lead shot forming the top of that heap, it is unlikely that oil will be able to do more than trickle upward.

In most cases, this interim solution will buy time while a relief well is drilled. The damaged well will then be sealed permanently by pumping cement or other dense material into the bottom of the damaged well. The presence of objects in the damaged well may complicate this sealing operation, but since upward flow is still present in the damaged well, the sealing fluids will be able to rise through the porous medium and produce the desired seal.

In summary, we propose to slow the flow of oil from a crippled oil well by filling that well with dense and probably streamlined objects. The heavy heap of objects, wedged against the well walls, will couple strongly to the rising oil, wasting its energy and extracting its upward momentum. With properly chosen iron or lead objects filling much of the volume of the well, the flow of oil will essentially stop.

We acknowledge the valuable insight, encouragement, and enthusiasm of Thomas F. Gallagher and Michael Fowler.

\bibliography{manuscript}

\end{document}